\documentclass[11pt,letterpaper]{article}
\usepackage{systeme} 
\usepackage[utf8]{inputenc}
\usepackage{multirow} 
\usepackage{booktabs} 
\usepackage[english]{babel}
\usepackage{amsmath}
\usepackage{amsfonts}
\usepackage{amssymb}
\usepackage{graphicx}
\usepackage[small,bf]{caption} 
\usepackage[left=3cm,right=3cm,top=2cm,bottom=3cm]{geometry}

\author{Jorge Pinochet}
\title{\textbf{Graphic relation between amplitude and sound intensity level}}
\begin{document}

\author{Jorge Pinochet$^{1}$, Walter Bussenius Cortada$^{2}$, Mauricio Sánchez Peña$^{3}$\\ \\
\small{$^{1}$\textit{Departamento de Física, Universidad Metropolitana de Ciencias de la Educación,}}\\
\small{\textit{Av. José Pedro Alessandri 774, Ñuñoa, Santiago, Chile.}}\\
\small{$^{2}$\textit{Facultad de Ciencias de la Educación, Universidad de Talca,}}\\
\small{\textit{Oriente 591, Linares, Maule, Chile.}}\\
\small{$^{3}$\textit{Facultad de Ingeniería, Universidad de Talca,}}\\ 
\small{\textit{Camino Los Niches Km. 1, Curicó, Chile.}}\\
\small{e-mail: jorge.pinochet@umce.cl}\\}

\date{}
\maketitle

\maketitle

\section{Introduction}

We present a simple experiment that allows us to demonstrate graphically that the intensity of sound waves is proportional to the square of their amplitude, a result that is theoretically analysed in any introductory wave course but rarely demonstrated empirically.\\

To achieve our goal, we use an audio signal generator that, when connected to a loudspeaker, produces sine waves that can be easily observed and measured using an oscilloscope. The measurements made with these instruments allow us to create a plot of amplitude versus sound intensity level, which verifies the mathematical relationship between amplitude and intensity mentioned above. Among the experimental errors, the plot obtained is in excellent agreement with what is theoretically expected. 

\section{Theoretical framework}

It is well known that the intensity level ($\beta$) of a sound wave is related to its intensity $I$ (in the international system it is measured in $W\cdot m^{-2}$) as [1]:

\begin{equation}
\beta = 10 log \left( \frac{I}{I_{0}} \right),
\end{equation}

where the factor 10 is due to the fact that the intensity level is expressed in decibels (one tenth of a bel) and the $I_{0}$ value corresponds to the lowest intensity that a human ear in good condition can detect ($10^{-12} W\cdot m^{-2}$).\\ 

\begin{figure}
  \centering
    \includegraphics[width=0.3\textwidth]{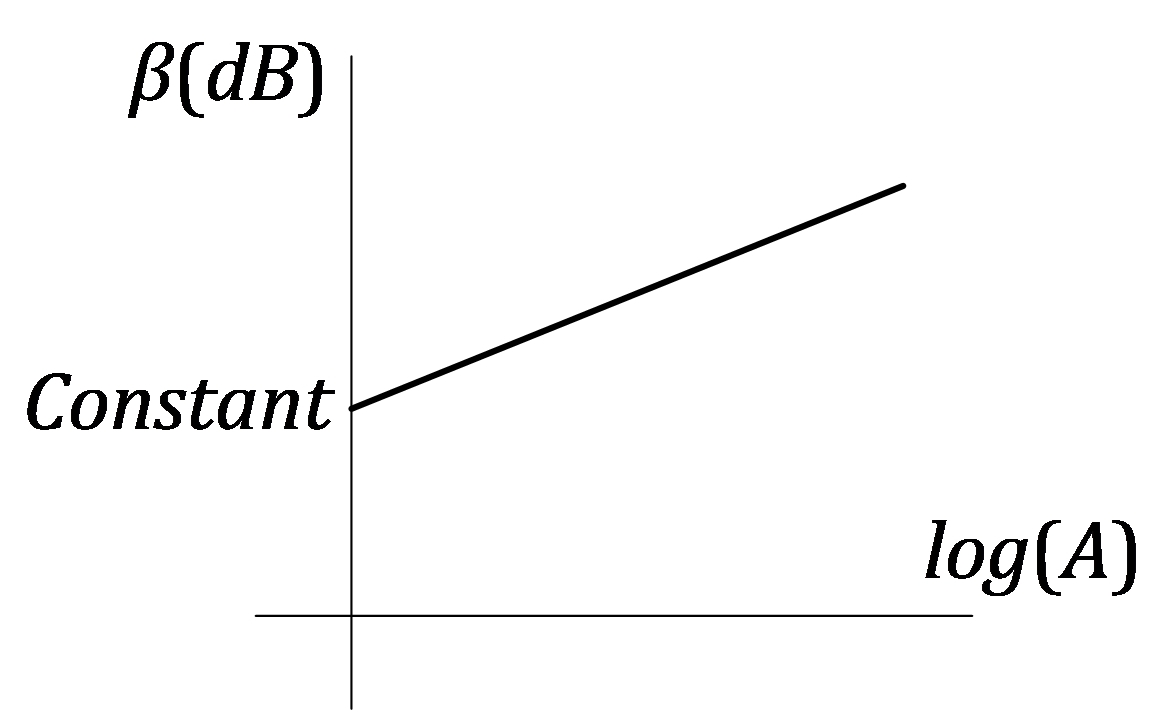}
  \caption{Expected theoretical plot of $\beta$ vs. $log(A)$.}
\end{figure}

As demonstrated in any general physics textbook, the intensity $I$ of a wave is proportional to the square of the amplitude $A$ [2]:

\begin{equation}
I = c_{0} A^{2},
\end{equation}

where $c_{0}$ is a constant. This means that the intensity of a wave increases faster than its amplitude. Starting from Eq. (2), Eq. (1) can be rewritten as:

\begin{equation}
\beta = 10 log \left( \frac{c_{0} A^{2}}{c_{1}} \right) 
\end{equation}

or,

\begin{equation}
\beta = 20 log \left( A \right) + c_{2},
\end{equation}

where $c_{1}$ and $c_{2}$ are constants. This equation shows us that if the intensity sound level, $\beta$ (dependent variable), is plotted versus the logarithm of its amplitude, $log (A)$ (independent variable), a line is obtained with a slope of 20, as illustrated in Fig. 1.\\

Using an audio signal generator to produce sound through a speaker, an oscilloscope to visualise and measure the amplitude of the emitted wave, and a sound level meter to measure the intensity level, it is possible to reproduce the plot in Fig. 1, which allows us to demonstrate experimentally the relationship between intensity and amplitude given by Eqs. (2) and (3). Fig. 2 schematises the experimental setup.

\begin{figure}[h]
  \centering
    \includegraphics[width=0.8\textwidth]{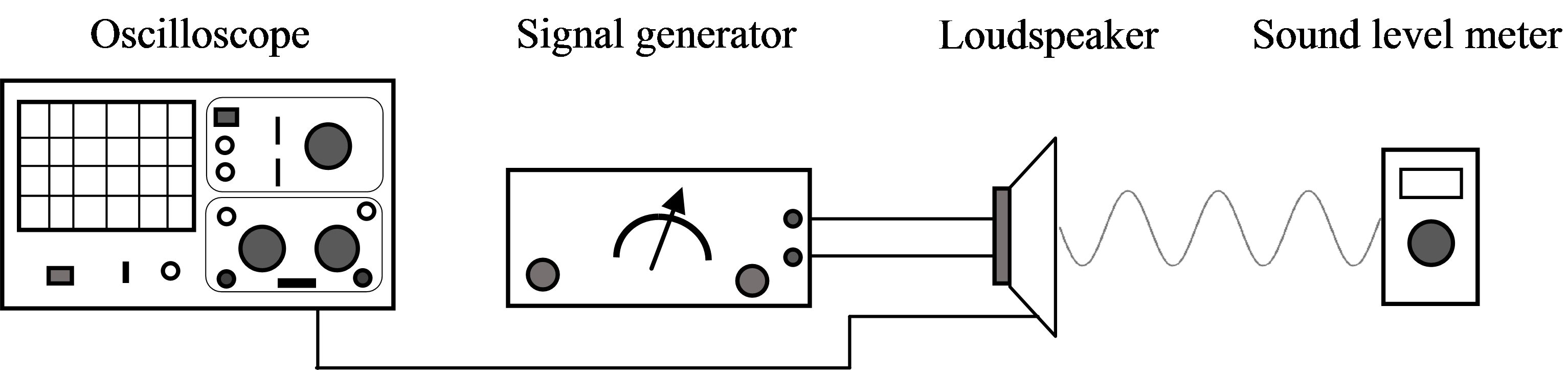}
  \caption{Schematic representation of the experimental setup.}
\end{figure}

\section{Materials list}

Audio signal generator EICO 377.\\		
Oscilloscope HAMEG HM 203-7.\\
Digital multimeter with a sound level meter MASTECH MS 8209.\\
Loudspeakers $3 W – 8 \Omega$.\\
Connection cables.\\
Coaxial cable.

\section{Graphic relation between amplitude and sound intensity level}

Before starting with the description of the experiment, it is important to note that for its correct performance, we recommend reducing the background noise to a minimum, ensuring that it is uniform during the experiment. For this reason, we also do not recommend taking measurements outdoors.\\

\begin{figure}[h]
  \centering
    \includegraphics[width=0.8\textwidth]{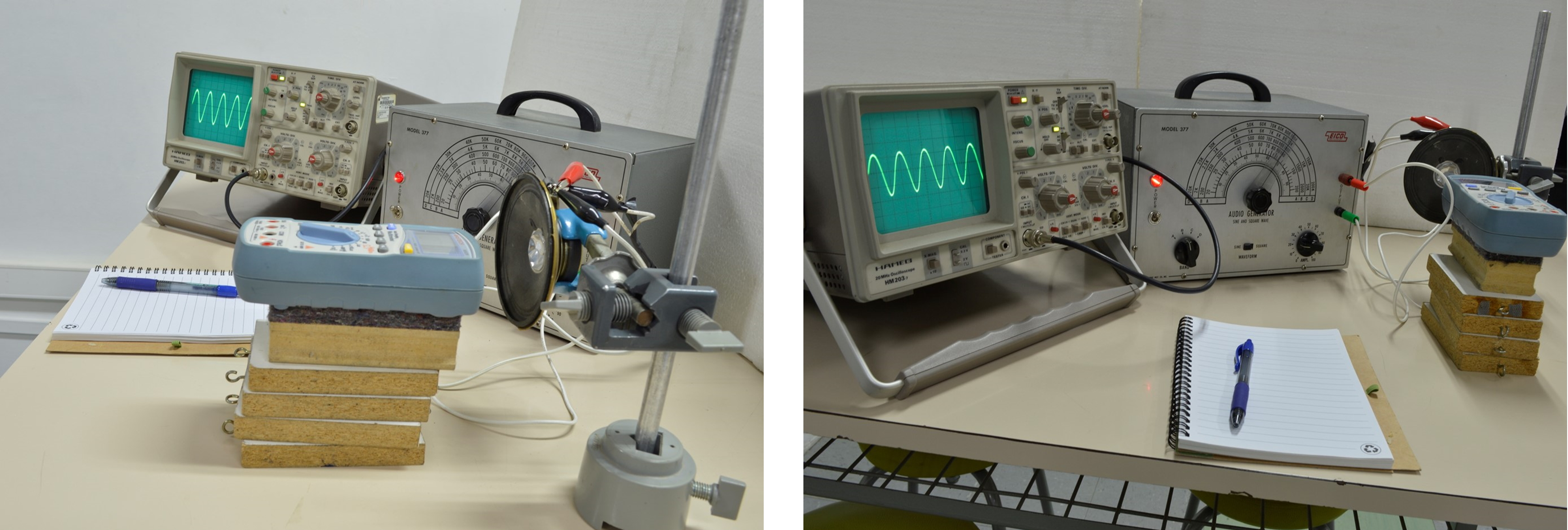}
  \caption{Experimental setup.}
\end{figure}

To carry out the experiment, a sinusoidal signal generator was connected to a speaker, which reproduces the pure sound emitted by the generator. In contrast, an oscilloscope was used to measure the amplitude of the generated wave, which is expressed in millivolts ($mV$). Finally, very close to the speaker, a sound level meter was installed to measure the intensity level due, on the one hand, to the background noise, and on the other, to the generated wave. Frequencies were established with the generator scale but were also measured with the oscilloscope.\\

\begin{figure}[h]
  \centering
    \includegraphics[width=0.9\textwidth]{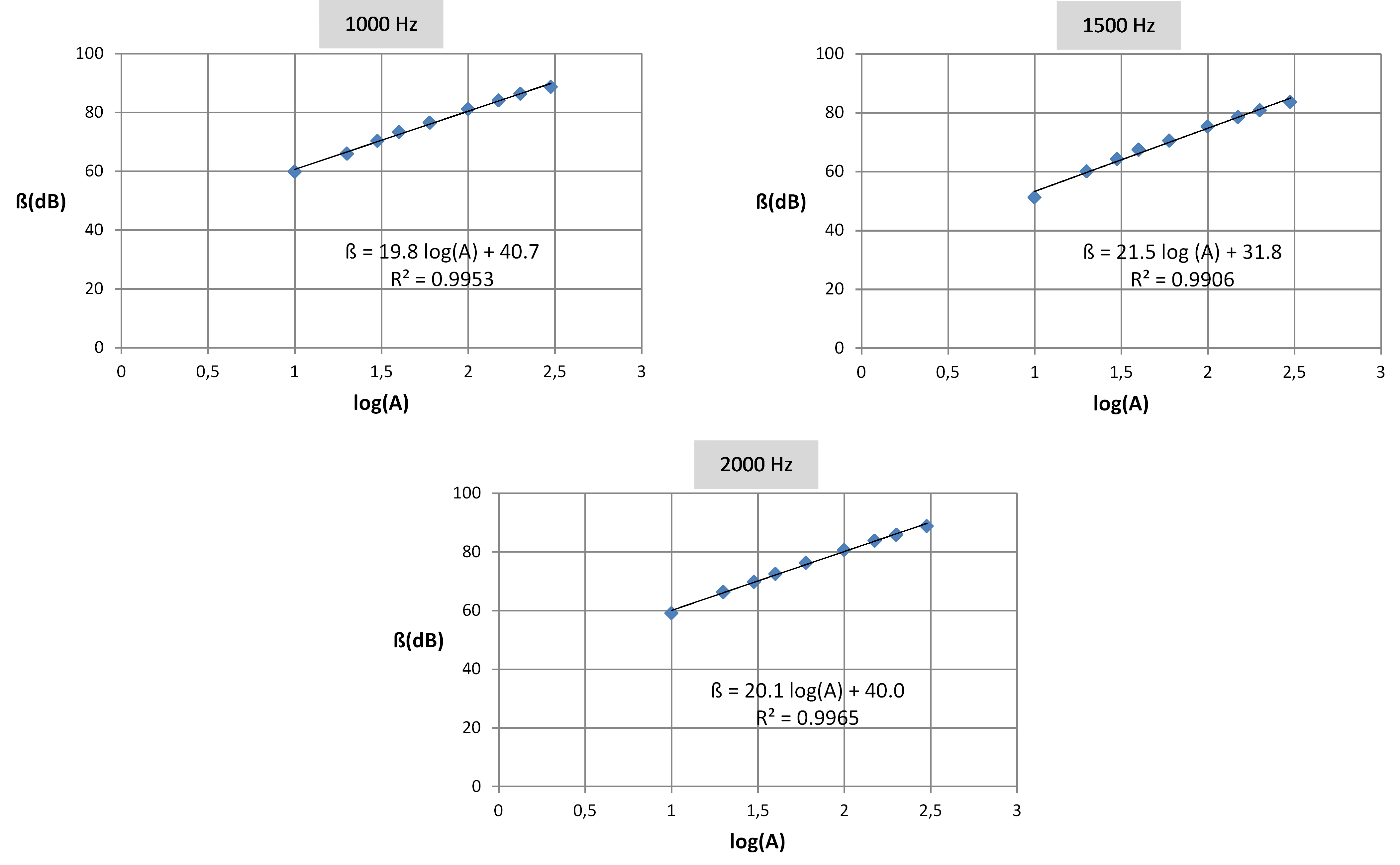}
  \caption{Sound intensity level vs. amplitude plots for three sets of measurements.}
\end{figure}

The experiment was carried out in a laboratory in the absence of students, with a background intensity level of  $33.3 dB$. Nine amplitude values were measured, which varied between 10 and 300 $mV$. \\

The experiment was tested with different frequencies and it was decided to perform it with those that gave the best results, i.e., those occurring in a range between 1000 and 2000 $Hz$. Higher frequencies (over 2000 $Hz$ and up to 5000 $Hz$) or lower (from 500 to 1000 $Hz$) caused a deviation from the linear relationship between sound intensity level and amplitude at the ends of the plotted line, and also generated a greater distortion of the sine wave shape, which is a consequence of the corresponding increase of amplitude.\\

The measurements at the intensity level ranged from 40 dB to a little less than $90 dB$, and the amplitude variations were in a range between 10 and 300 $mV$; therefore, both the amplitude variation range and the range intensity level were significant. Fig. 3 shows the experimental setup.\\

Table 1 summarises the measurements for frequencies of 1000, 1500 and 2000 $Hz$. For each of these frequencies, the table shows the amplitude $A$ (in $mV$), the sound intensity level $\beta$ (in $dB$) and the logarithm of the amplitude $log (A)$. \\ 

The plots for the three sets of measurements are shown in Fig. 4. As can be seen, the linear trend is satisfactory. Furthermore, in all three cases, the experimental slope ($19.8\pm0.5$, $21,5\pm0.8$, and $20,1\pm0.5$) is very close to the expected theoretical value (20). \\ 

\begin{table}[htbp]
\begin{center}
\caption{Summary of amplitude and intensity level measurements.}
\begin{tabular}{l l l l l l l l l} 
\toprule
\multicolumn{3}{l}{First set of measures} &
\multicolumn{3}{l}{Second set of measures} & 
\multicolumn{3}{l}{Third set of measures} \\
\multicolumn{3}{l}{(1000 Hz)} &
\multicolumn{3}{l}{(1500 Hz)} & 
\multicolumn{3}{l}{(2000 Hz)} \\
\midrule
$A (mV)$ & $\beta (dB)$ & $log (A)$ & $A (mV)$ & $\beta (dB)$ & $log (A)$ & $A (mV)$ & $\beta (dB)$ & $log (A)$ \\
$\pm 5\%$ & $\pm 0.3\%$ &  & $\pm 5\%$ & $\pm 0.3\%$ &  & $\pm 5\%$ & $\pm 0.3\%$ &  \\ 
\midrule
10 & 59.8 & 1.000 & 10 & 51.3 & 1.000 & 10 & 59.1 & 1.000 \\ 
\midrule
20 & 66.0 & 1.301 & 20 & 60.1 & 1.301 & 20 & 66.3 & 1.301 \\
\midrule
30 & 70.3 & 1.477 & 30 & 64.3 & 1.477 & 30 & 69.7 & 1.477 \\
\midrule
40 & 73.3 & 1.602 & 40 & 67.4 & 1.602 & 40 & 72.5 & 1.602 \\
\midrule
60 & 76.5 & 1.778 & 60 & 70.5 & 1.778 & 60 & 76.3 & 1.778 \\
\midrule
100 & 81.0 & 2.000 & 100 & 75.3 & 2.000 & 100 & 80.7 & 2.000 \\
\midrule
150 & 84.1 & 2.176 & 150 & 78.5 & 2.176 & 150 & 83.8 & 2.176 \\
\midrule
200 & 86.3 & 2.301 & 200 & 80.9 & 2.301 & 200 & 85.9 & 2.301 \\
\midrule
300 & 88.7 & 2.477 & 300 & 83.7 & 2.477 & 300 & 88.8 & 2.477 \\
\bottomrule
\end{tabular}
\label{Summary of amplitude and intensity level measurements}
\end{center}
\end{table}

The uncertainties were estimated using the least squares method. For this reason we have included the $R$ squared coefficient in the graphs. The uncertainties are basically associated with the amplitude of the wave displayed by the oscilloscope. We estimate its value between 3 and 5\% depending on the scale of the instrument (considering half of the smallest division of the oscilloscope). The uncertainty associated with the sound level meter is lower than that of the oscilloscope (except for very small amplitudes), and is in the range $0.1 - 0.2 dB$, approximately 0.3\% (on the order of 10 times lower than the uncertainty associated with the amplitude). Therefore, it is feasible to neglect this last value.\\ 

It was not possible to extend the measurements beyond the range $10 - 300 mV$. The reason is that for values less than $10 mV$ the uncertainty in the oscilloscope reading increases due to the widening of the wave (see Fig. 5, left), while for values greater than $300 mV$ the wave deforms significantly (see Fig. 5, right). 

\begin{figure}
  \centering
    \includegraphics[width=0.8\textwidth]{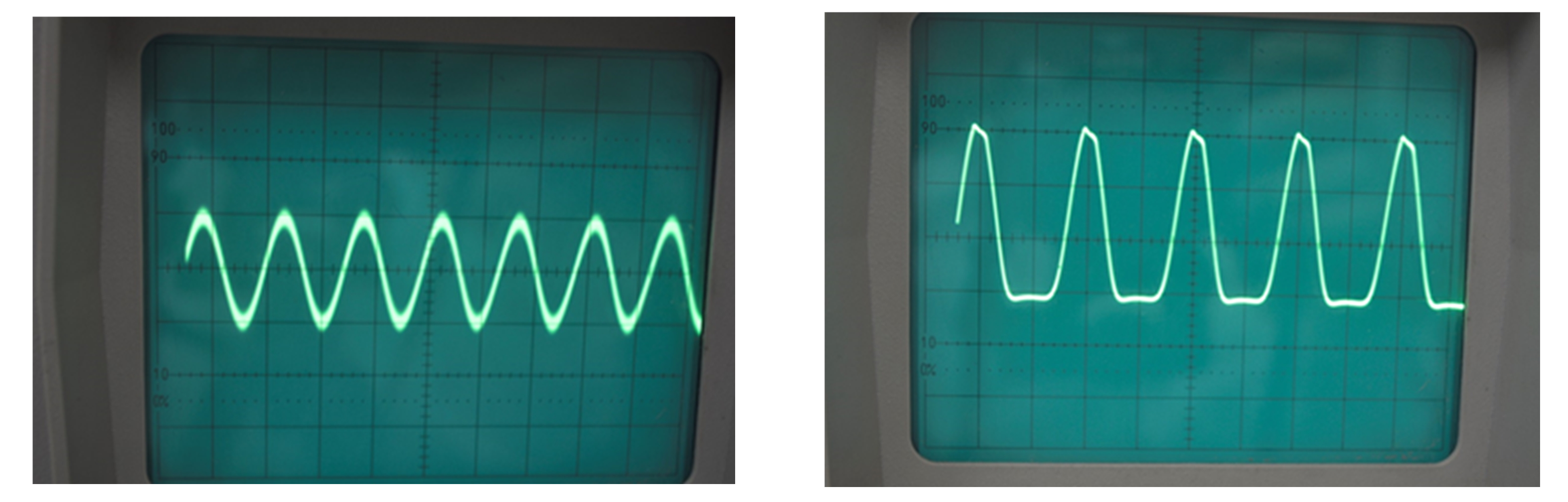}
  \caption{ Distortion in the waveform displayed by the oscilloscope.}
\end{figure}

\section{Comments and conclusions}
Firstly, it seems remarkable to us that with accessible materials and with a simple experimental setup, it is possible to establish the relationship between intensity and amplitude, a relationship that is generally only addressed theoretically in wave courses. In particular, the excellent agreement between the expected and obtained slopes is gratifying. Another aspect that should be highlighted is that, together with the graphic representation of the results, the experiment allows us to visualise the waves and their amplitude variations in the oscilloscope, and in parallel it makes it possible to measure the intensity level and perceive it with the ear, that is, it makes it possible to appreciate the concepts intuitively and concretely.\\ 

In relation to the uncertainties associated with the experiment, we raise three aspects. The first is that the amplitude is visualised at a guess on the oscilloscope screen, which introduces uncertainty in the value of the amplitude. The second aspect is that the intensity level measurement with the sound level meter is affected by background noise that cannot be controlled or eliminated, which generates fluctuations in the intensity level. Finally, with respect to Fig. 4, it is important to note that the difference between the intercept of the first graph (40.7) with respect to the other two (31.8 and 40), is due to the point of cut with the vertical axis, which is determined by the sensitivity of the speaker and cannot be controlled. In relation to this issue, it is important to note that to carry out the experiment we used a low-cost speaker that was extracted from a broken radio. Taking into account all the antecedents, the estimated fluctuations for the slope are of the order of 5\%, and since its value is 20, the estimated variation is of the order of unity.\\ 

It is important to bear in mind that once the experiment has been set up, the measurements and their respective graphic representation require a time of $\sim 15 min$, meaning it is perfectly feasible to carry out this experimental activity in a demonstrative way in a classroom (with silence from the students).\\ 

A key aspect that we believe is important for the teacher to discuss with their students is that although a sound wave is generated by pressure variations in the air, and therefore the amplitude of a wave of this type should be associated with said variations, in the experiment, we use an amplitude scale expressed in $mV$. As we know, the reason is that the potential difference applied to the speaker was measured, and therefore the corresponding amplitude is expressed in $mV$. However, said voltage is related to the current in the speaker, which in turn is related to the amplitude of the movement generated in the speaker and the pressure differences of the sound wave.\\ 

In summary, considering the simplicity of the presented experiment and the accuracy of the results obtained, we hope that this work constitutes a contribution to the work carried out by teachers who teach wave courses, both at school and university level.

\section*{Acknowledgments}
I would like to thank to Daniela Balieiro for their valuable comments in the writing of this paper. 

\section*{References}

[1] R.A. Serway, J.W. Jewett, Physics for Scientists and Engineers with Modern Physics, 8 ed., (Thomson, Cengage Learning), 2010, p. 480.

\vspace{2mm}

[2] P.A. Tipler, Physics for Scientists and Engineers, 5 ed., (W. H. Freeman and Company, New York, 2004), p. 476.

\end{document}